\documentclass[aps,pra,twocolumn,epsf]{revtex4}

\usepackage{amsmath,float,graphicx,subfigure}

\begin{document}  
 
\title{Photonic spectrum of bichromatic optical lattices}
\author{Stefan Rist$^1$ Patrizia Vignolo$^2$, and Giovanna
Morigi$^1$} \affiliation{$^1$ Grup d'{\`O}ptica, Departament de F{\'i}sica,
Universitat Aut\`onoma de Barcelona, 08193 Bellaterra, Spain
\\$^2$ Institut Non Lin\'eaire de Nice,
Universit\'e de Nice-Sophia Antipolis, CNRS, 1361 route des 
Lucioles, 
06560 Valbonne, France}

\begin{abstract} We study the photonic spectrum of a one-dimensional
optical lattice possessing a double primitive cell, when the atoms are well localized at the lattice minima.
While a one-dimensional lattice with a simple Wigner-Seitz cell always possesses a photonic bandgap at the atomic resonance,
in this configuration the photonic transmission spectrum may exhibit
none, double or multiple photonic bandgaps depending on the ratio between the interparticle distance $\varrho$
inside the cell and the cell size $a$. The transmission spectra of a
weak incident probe are evaluated when the atoms are trapped in free space and inside an optical resonator for realistic experimental parameters.  \end{abstract}

\maketitle

\section{\label{sec:intro}Introduction}

Ultracold atoms in optical lattices constitute a paradigmatic
system, which allows one control over several parameters, thereby
mimicking dynamics typical of condensed matter systems~\cite{Bloch-Review,Maciej-Review}. A
remarkable feature of optical lattices is that the bulk
periodicity is here controlled by engineering the geometry of the
propagating beams, which determine the light potentials. Differing
from ordinary crystals in condensed matter, the size of the Wigner-Seitz cell is of the order of the light wavelength.
One consequence is that the light, coupling with the atomic transitions, is also diffracted by the crystallyne structure which the atoms form~\cite{Grynberg}.

It has been observed that the modulation of the atomic density in these systems,
and hence of the refractive index, makes optical lattices a photonic bandgap
material~\cite{Deutsch1995a}. Theoretical works studied the photonic bandgap for one-dimensional and
three-dimensional atomic structures~\cite{Deutsch1995a,Lagendijk,Lambropoulos-Review,Artoni,Pritchard,Carusotto}.
Besides potential applications for nonlinear optics, the full understanding of these properties is important in order to measure the quantum state of cold atoms using light~\cite{Mekhov}, thereby opening interesting avenues in quantum information processing for implementing photonic interfaces using ultracold atoms~\cite{Zoller-roadmap}.

In this work we study theoretically the photonic properties of biperiodic
optical lattices, in a setup similar to the ones realized experimentally in~\cite{Davidson2006,Fallani,Barmettler2008a}. We focus on a one-dimensional configuration, and develop a full quantum model for the light-matter interactions, assuming that the atoms are well localized at the lattice minima. The photonic spectra and the probe transmission are evaluated when the optical lattice is in free space and inside a standing wave optical resonator, as a function of the interparticle distance $\varrho$ inside the primitve cell.

This article is organized as follows. In Sec.~\ref{Sec:1} the theoretical model is described
and the basic approximations are introduced. In Sec.~\ref{Sec:2}
the photonic spectra are reported and discussed, and the transmission spectra for a weak probe are evaluated in Sec.~\ref{Sec:3}. In
Sec.~\ref{Sec:4} the photonic properties of bichromatic optical lattices inside a resonator are analyzed. The conclusions are presented
in Sec.~\ref{Sec:5}.

\section{Theoretical model} \label{Sec:1}

The physical system we consider is a one-dimensional periodic
distribution of atoms in a light potential with a double primitive
cell. We assume a sequence of $N$ atoms of mass $m$ in a
standing wave created with lasers along the $x$ direction. We denote the atomic positions
along $x$ by $x_j$, with $j=1,\ldots, N$. Denoting
by $a$ the size of the Wigner-Seitz cell, the positions are given
by
\begin{eqnarray} x_j&=&\ell a~~\mbox{for}~j=2\ell\,,\nonumber\\
&=&\ell a+\varrho~~\mbox{for}~j=2\ell+1\,,\nonumber \end{eqnarray}
where $\ell=0,1,\ldots,M-1$, and $M=N/2$ is the number of cells
(assuming $N$ even for convenience). In the case here discussed,
we set the size $a=\lambda$, where $\lambda$ is the wavelength of the
light which interacts with the dipolar transitions of the atoms.

\begin{figure} \begin{center}
\includegraphics[width=0.9\linewidth,clip=true]{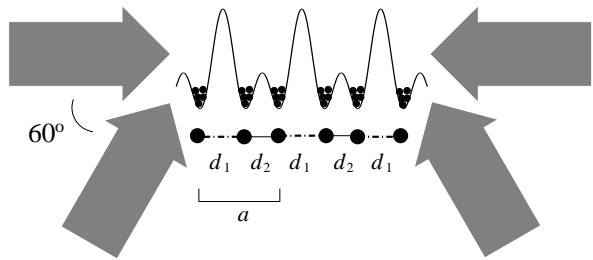}
\caption{A possible optical realization of the double period
1D lattice, considered in this work, can be obtained by using a monochromatic standing wave
with wavelength $\lambda$, to which two laser beams are
superposed, such that they are rotated by angles of 60$^{\circ}$
and 120$^{\circ}$ with respect to the axis of the lattice. Here, $d_1=\varrho$, $d_1+d_2=\lambda$ and the size of the Wigner-Seitz cell is $a=\lambda$.
} \label{lattice} \end{center} \end{figure}

Such configuration can be experimentally realized by using a monochromatic standing wave
with wavelength $\lambda$, to which two laser beams are
superposed, such that they are rotated by angles of 60$^{\circ}$
and 120$^{\circ}$ with respect to the axis of the lattice, as
shown in Fig. \ref{lattice}. Upon setting the relative phases, the
resulting potential for the atoms has the form \begin{equation}
U(x)\propto\beta^2\cos^2(kx/2)+\cos^2(kx) \end{equation} and the
distances between adjacent wells are
$d_1=\lambda(1-\frac{1}{\pi}\,{\rm acos}\,\frac{-\beta^2}{4})$ and
$d_2=\frac{\lambda}{\pi}\,{\rm acos}\,\frac{-\beta^2}{4}$, with
$d_1+d_2=\lambda.$
Another possible realization is found by superposing two laser beams along the $x$-axis, with a half frequency \cite{Folling2007} or
with a three fourth frequency
respect to the frequency of the main lattice \cite{Fallani2007}. Upon setting the relative phases, the four-atomic elementary cell has the structure $d_1-d_1-d_2-d_2$ (with $d_1+d_2=\lambda$) and the crystal has essentially the same
spectral properties than the biatomic one considered in this paper. In this work we will also consider biperiodic lattices, where the  atoms composing the Wigner-Seitz cell may have different scattering properties, for instance, they can belong to different species or
belong to the same species but are prepared in different hyperfine states. Under the assumption that the frequencies of the two
transitions are sufficiently close to allow significant coupling
with the same probe, such lattices could be realized with linearly polarized counterpropagating beams, controlling the angle between the polarization~\cite{Bloch-lattice}.

In developing the theoretical model we
will make the following assumptions: {\it (i)} The atoms are well
localized at the lattice points, and the size of the atomic wave
packet is very small with respect to the laser wavelength
(Lamb-Dicke regime~\cite{LambDicke}). This situation can be
realized when the atoms are deep in the Mott-insulator quantum
state of the biperiodic potential~\cite{Bloch-Review}. We will
treat the atoms as pointlike, hence considering the response of
the medium at lowest order in the expansion of the size of the
wave packet over the wavelength. {\it (ii)} We consider the coupling of the
lattice only with modes of the electromagnetic field which
propagate along the lattice. This approximation is valid when the atoms are placed, for instance, inside a bad cavity with sufficiently large cooperativity~\cite{Kimble_Review} or a hollow-core fiber~\cite{Chang,Ketterle}.

\subsection{Hamiltonian}

The Hamiltonian $H$ for the total system, composed by the atomic
spins aligned along the $\hat{x}$ axis, and the modes of the
electromagnetic field propagating along $\hat{x}$, reads
$$H=H_{\rm dip}+H_{\rm emf}+H_{\rm int}\,.$$ Here, \begin{eqnarray}
\label{H:dip} H_{\rm
dip}=\sum_j\hbar\omega_j\sigma_j^{\dagger}\sigma_j \end{eqnarray}
describes the array of dipoles, with $j=1,\ldots,N$ labeling the
atoms, where $\sigma_j$ is the dipolar lowering operator and
$\sigma_j^{\dagger}$ its adjoint, whereby the relevant transitions of the atoms at the even
(odd) sites, $j=2\ell$ ($j=2\ell +1$), have dipole moments
$\mathcal D_1$ ($\mathcal D_2$) and transition frequency
$\omega_1$ ($\omega_2$). The Hamiltonian for the modes of the
electromagnetic field takes the form \begin{eqnarray} H_{\rm
emf}=\sum_{k}\sum_{n=1,2} \hbar\omega_ka_k^{(n)\dagger}a_k^{(n)}\,,
\end{eqnarray} where the operators $a_k^{(n)}$ and
$a_k^{(n)\dagger}$ annihilate and create a photon in the
electromagnetic field mode at frequency $\omega_k$, wave-vector
$k=\omega_k/c$, with $c$ light velocity, and polarization
$\hat{\epsilon}_n=\hat{\epsilon}_1, \hat{\epsilon}_2$, with
$\hat{\epsilon}_1\perp \hat{\epsilon}_2\perp \hat{x}$. Finally,
the interaction between photons and atoms is described by the
Hamiltonian term (in Coulomb gauge) \begin{equation} H_{\rm
int}=-e \sum_j \frac{p_j}{m_j}\cdot A_{\perp}(x_j)\,, \end{equation}
where $m_j$ is the mass of the atom $j$, $p_j$ is the momentum
operator of the electron at atom $j$, \begin{equation} p_j={\rm i}
\frac{m_j}{e} \omega_j D_j (\sigma_j^{\dagger}-\sigma_j)\,,
\end{equation} and
$A_{\perp}(r)$ is the transverse vector potential \begin{equation}
A_{\perp}(r)=\sum_k\sum_{n=1,2}
\sqrt{\frac{\hbar}{2\varepsilon_0\omega_k
V}}(a_k^{(n)}\hat{\epsilon}_n {\rm e}^{{\rm i}k\cdot r} + {\rm
H.c.})\,, \end{equation} with $\varepsilon_0$ the vacuum
permittivity and $V$ the quantization volume. Note that we have used a plane waves decomposition,
applying periodic boundary conditions at the lattice borders.

\subsection{Weak excitation regime}

In this work we consider that the atomic transitions are driven
well below saturation, and correspondingly the mean number of
photonic excitations inside the system is much smaller than the
total number of spins $N$. In this regime we use the
Holstein-Primakoff representation of spin operators~\cite{Holstein}, and expand
all operators at the lowest orders in the powers of bosonic operators $b_j$,
\begin{eqnarray} \sigma_j^{\dagger} &=& b_j^\dagger\,
(1-b_j^\dagger b_j)^{1/2}\simeq b_j^{\dagger}\left(1-b_j^\dagger
b_j/2\right)\,,
\\
\sigma_j^- &=& (1-b_j^\dagger b_j)^{1/2}\,  b_j
\simeq \left(1-b_j^\dagger b_j/2\right) b_j\,,\\
\sigma_j^z &=& -\frac{1}{2} + b_j^\dagger b_j\,. \end{eqnarray}
In this representation, the Hamiltonian for
the dipoles becomes the sum of $N$ harmonic oscillators,
\begin{equation} H_{\rm dip}=\sum_j \hbar\omega_jb_j^\dagger b_j
\end{equation} where we discarded the constant term. The
interaction term reads (in the Rotating Wave Approximation) \begin{equation} H_{\rm int}=
H^{(1)}+H^{(3)}\,, \end{equation} with \begin{eqnarray}
&&H^{(1)}=
\sum_{j,k,n}\hbar\mathcal G_{j,k}^{(n)}b_j^{\dagger}a_k^{(n)}~{\rm e}^{{\rm i}kx_j}+{\rm H.c.}\,,
\end{eqnarray}
while $H^{(3)}$ describes the corrections beyond the linear response.
Here, $\mathcal G_{j,k}^{(n)}$ is the coupling
strength of the atom $j$ with the mode $(k,n)$ and is given by
 \begin{equation} \mathcal G_{j,k}^{(n)}=-{\rm i}\omega_j{\mathcal D}_j\cdot
\hat{\epsilon}_n\sqrt{\frac{1}{2V\varepsilon_0\hbar
\omega_k}}. \end{equation} We will consider the limit in which we
can truncate the expansion and approximate $H_{\rm int}\approx
H^{(1)}$, thereby restricting to the case in which the medium
polarization is linear in the electric field amplitude.

\subsection{Spin waves}

Given the periodic structure, it is convenient to describe the
dipolar excitations in momentum space. At this purpose, for a
sufficiently large crystal we assume Born-von Karman periodic
boundary conditions, and consider the spin-wave excitations
\begin{eqnarray}
&&b_q=\frac{1}{\sqrt{M}}\sum_{\ell=0}^{M-1}b_{2\ell}{\rm
e}^{-{\rm i}\ell q a}\,,\\
&&{d}_q=\frac{1}{\sqrt{M}}{\rm e}^{-{\rm i}q\varrho}
\sum_{\ell=0}^{M-1}b_{2\ell+1}{\rm e}^{-{\rm i}\ell q a}\,,
\end{eqnarray} with $q$ the wave vector sweeping the first
Brillouin zone (BZ).
We denote by $$G_0=2\pi/a$$ the elementary vector of the reciprocal
lattice, such that the interval of the first BZ is
$[-G_0/2,G_0/2]$. Using the relation $\sum_{\ell=0}^{M-1}\exp({\rm
i}(q-q^{\prime})\ell a)=M\delta_{qq^{\prime}}$ where the equality
$q=q^{\prime}$ is defined modulus a vector $G$ of the reciprocal
lattice, the Hamiltonian terms transform as \begin{eqnarray}
&&H_{\rm dip}
=\sum_{q\in BZ} \hbar\left(\omega_1 b_q^{\dagger}b_q+\omega_2d_q^{\dagger}d_q\right)\,,\\
&&H^{(1)}= \sum_{G,n}\sum_{q\in BZ}\hbar\sqrt{M} \left(\mathcal
G_{1,q+G}^{(n)}b_q^{\dagger}+{\rm e}^{{\rm i}G\varrho} \mathcal
G_{2,q+G}^{(n)}d_q^{\dagger}\right) a_{q+G}^{(n)} \nonumber \\ & &
+{\rm H.c.}\,, \nonumber\\\end{eqnarray} where the
quasi-momentum verifies the relation $k=q+G$. In this
form, the Hamiltonian can be rewritten as the sum of $M$
Hamiltonian terms, $H=\sum_{q\in BZ} H_q$, where \begin{eqnarray}
H_q&=&\hbar\omega_1
b_q^{\dagger}b_q+\hbar\omega_2d_q^{\dagger}d_q+
\hbar\sum_{G,n}\omega_{q+G}a_{q+G}^{\dagger(n)}a_{q+G}^{(n)} \nonumber \\
&+&\hbar \sum_{G,n}\left[\sqrt{M} \left(\mathcal
G_{1,q+G}^{(n)}b_q^{\dagger}+{\rm e}^{{\rm i}G\varrho} \mathcal
G_{2,q+G}^{(n)}d_q^{\dagger}\right) a_{q+G}^{(n)}+{\rm
H.c.}\right]. \nonumber \\ \end{eqnarray} This separation is only
valid in the linear regime since saturation effects, described by $H^{(3)}$, mix the manifolds identified by the
Hamiltonian terms $H_q$.

\section{The photonic band structure} \label{Sec:2}

In this section we study the photonic spectrum of the biperiodic
structure assuming that the polarization of the incident light,
say $\epsilon_1$, is parallel to the dipole moments $\mathcal D_1$ and
$\mathcal D_2$. Hence, we drop the polarization superscripts where they
appear.
The photonic band structure is found by solving the Heisenberg
equations of motion for each Hamiltonian block $H_q$,
\begin{eqnarray} \label{eqn:bandstructure} &&\dot{a}_{q+G}=-{\rm
i}\omega_{q+G} a_{q+G} \nonumber \\
& & -{\rm i} \sqrt{M} \left( \mathcal G^*_{1,q+G}b_q-{\rm e}^{-{\rm i}G\varrho} \mathcal G^*_{2,q+G}d_q \right )\,,\\&&\nonumber\\
&&\dot{b}_q=-{\rm i}\omega_1 b_q-{\rm
i}\sqrt{M}\sum_{G} \mathcal G_{1,q+G} a_{q+G}\,,\\
&&\dot{d}_q=-{\rm i}\omega_2 d_q-{\rm i} \sqrt{M}\sum_{G} {\rm
e}^{{\rm i}G\varrho} \mathcal G_{2,q+G} a_{q+G}\,, \end{eqnarray}
where $\omega_q=c |q|$. Hence, the spin wave at wavevector $q$
couples in principle with all photonic modes at wavevectors $q+G$.
Nevertheless, only the coupling of the atomic transition with the
quasi-resonant modes at wavevectors $Q = \pm G_0 $ is significant.
Taking into account only the relevant coupling, we can solve
analytically the eigenvalue problem around $q\simeq 0$ in the BZ
and assuming $\omega_1=\omega_2$. In this limit one finds four
eigenfrequencies, 
\begin{widetext} \begin{eqnarray}
\label{eqn:bandgap} \nu_{j,\pm}=\frac{\omega_Q+\omega_1}{2} \pm
\sqrt{\left(\frac{\omega_Q-\omega_1}{2}\right )^2+M \mathcal
G^2\left(1-(-1)^j \sqrt{1-\left(\frac{2|\mathcal G_{1,Q}\mathcal
G_{2,Q}|}{\mathcal G^2}\right)^2 \sin^2 G_0\varrho}\right )}\,,
\end{eqnarray} \end{widetext} 
where $\mathcal G=\sqrt{|\mathcal
G_{1,Q}|^2+|\mathcal G_{2,Q}|^2}$ and $j=1,2$. They determine the
edges of two photonic bandgaps, one at the frequencies between
$\nu_{1,-}$ and $\nu_{2,-}$ and the second between $\nu_{2,+}$ and
$\nu_{1,+}$. We note that the bandgap size depends on the
interparticle distance $\varrho$ inside the Wigner-Seitz cell but
is independent on the number of cells $M$, and thus it is constant in the thermodynamic limit: in fact the
quantization volume $V\propto 1/\sqrt{M}$ (in 1D) gives that $\mathcal
G\propto 1/\sqrt{M}$, so that the dependence on $M$ in
Eq.~(\ref{eqn:bandgap}) cancels out. 

For $\varrho\to 0$, in the
limit of the monoperiodic array, one finds a single bandgap with
size $\Delta\omega=\nu_{1,+}-\nu_{1,-}$. 
For $\varrho>0$
($\varrho<a$) this interval is reduced: a frequency window opens
inside the gap, where light is transmitted, and whose width is
given by $\Delta\omega_{\varrho}=\nu_{2,+}-\nu_{2,-}$. The size of
the two bandgaps is minimum at $\varrho=a/4$, and it vanishes at
this point when $\mathcal G_{1,Q}=\mathcal G_{2,Q}$. In this
specific case, hence, the lattice becomes completely transparent.
This is simply understood, considering that the bandgap results
from an interference effect due to multiple scattering by all
atomic planes, and it hence depends on the phase relations between
the fields scattered by each plane. For this specific
configuration, where $\varrho=\lambda/4,3\lambda/4$, the phase
accumulated due to scattering of the first atom of the cell
cancels out with the phase due to scattering by the second atom.
As a result, the total phase accumulated from scattering with the
two atoms of the cell is zero, and the medium hence behaves as it
were completely transparent. Note that a monoperiodic array is also found 
for $\varrho=a/2$, when all atoms have the same scattering properties. In this case the periodicity is halved, and the 
first BZ doubles, $[-G_0,G_0]$. The photonic spectra of this system have been discussed in~\cite{Pritchard}.

\begin{figure*}[htp] \centering \subfigure[]{
\includegraphics[width=.4\textwidth]{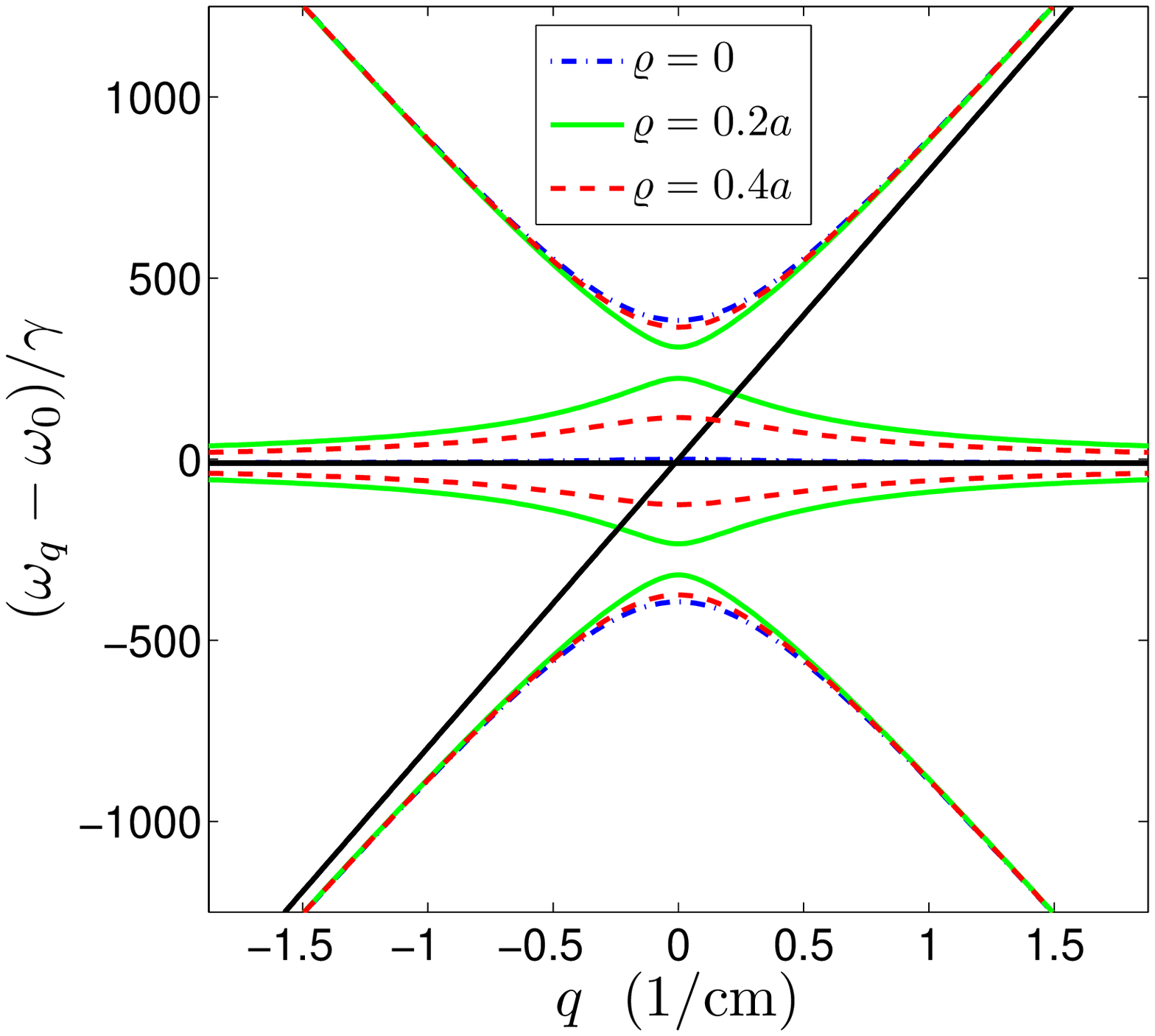}} \subfigure[]{
\includegraphics[width=.4\textwidth]{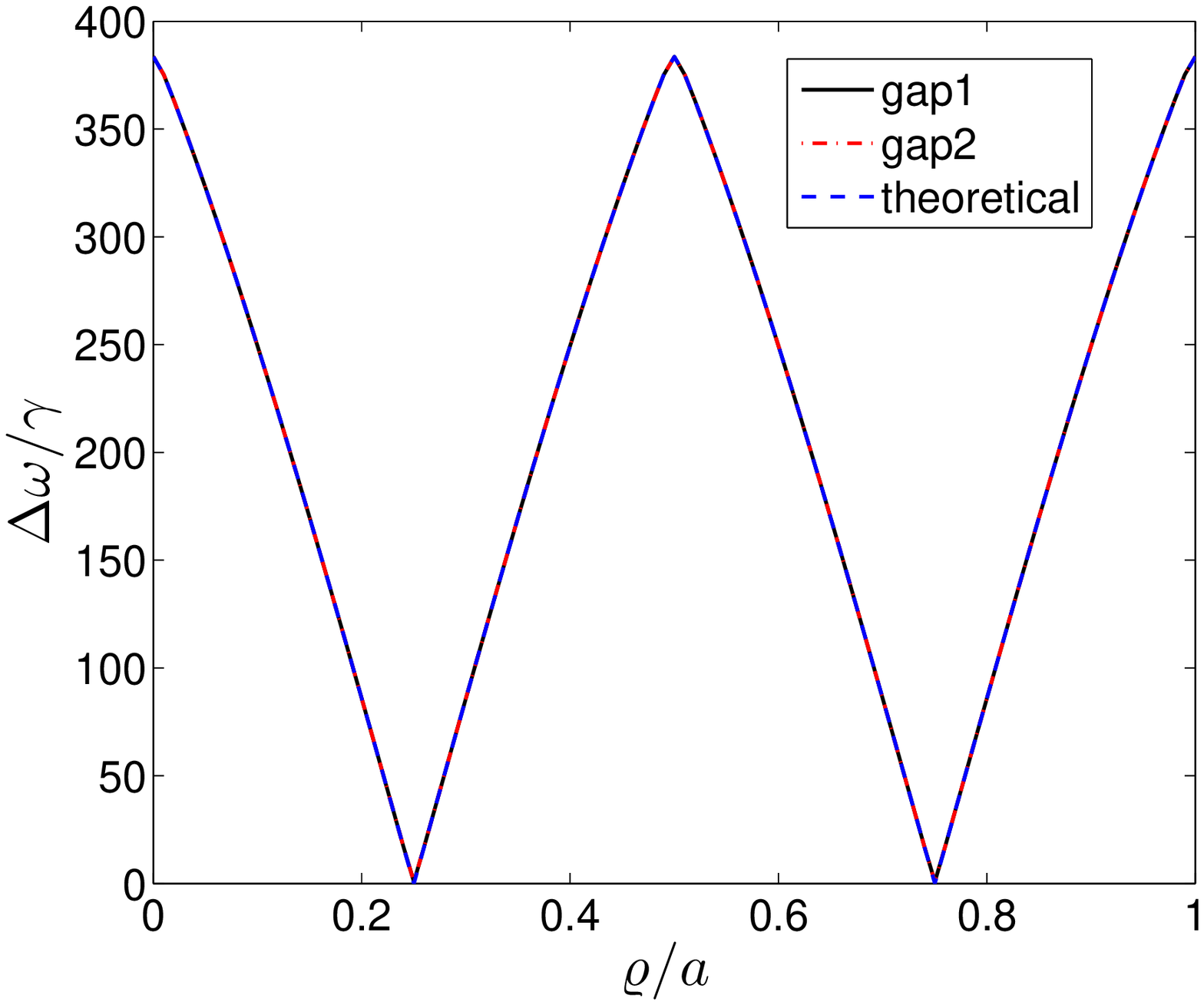}}
\vspace{.1in} \caption{\label{Fig:Spectrum:1} (Color online) (a) Polariton
dispersion for a 1D bichromatic lattice for different values of
$\varrho$ obtained by summing over $40$ BZ, when the atoms are trapped
in a hollow fiber, whose fundamental mode is gaussian with waist $w=5\mu {\rm m}$ 
\cite{Ketterle}. The curves are evaluated for
$\omega_1=\omega_2=\omega_0-10\gamma$, considering the D2-line of
$^{85}$Rb atoms, with $\lambda=780$ nm and $\gamma=2\pi\times 6$
MHz. The solid black line is plotted for comparison, and
corresponds to the case in which atoms and field are not coupled ($\mathcal
G_j=0$). (b) Photonic bandgap (numbered from lower to higher
frequency) as a function of $\varrho/a$, compared with the
analytical prediction obtained from Eq. (\ref{eqn:bandgap}). }
\end{figure*}

\begin{figure*} \subfigure[]{
\includegraphics[width=.4\textwidth]{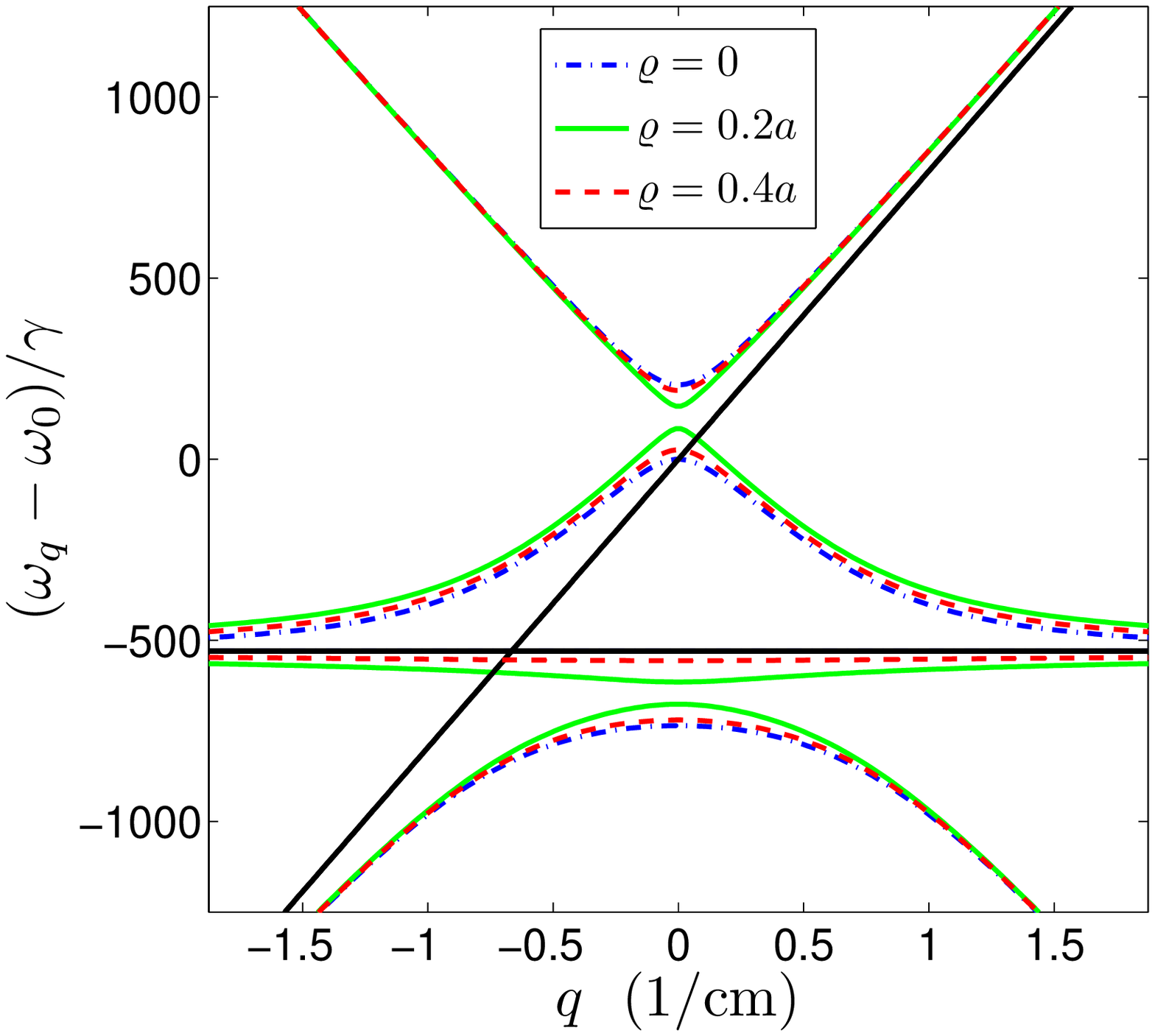}}
\vspace{.1in} \subfigure[]{
\includegraphics[width=.4\textwidth]{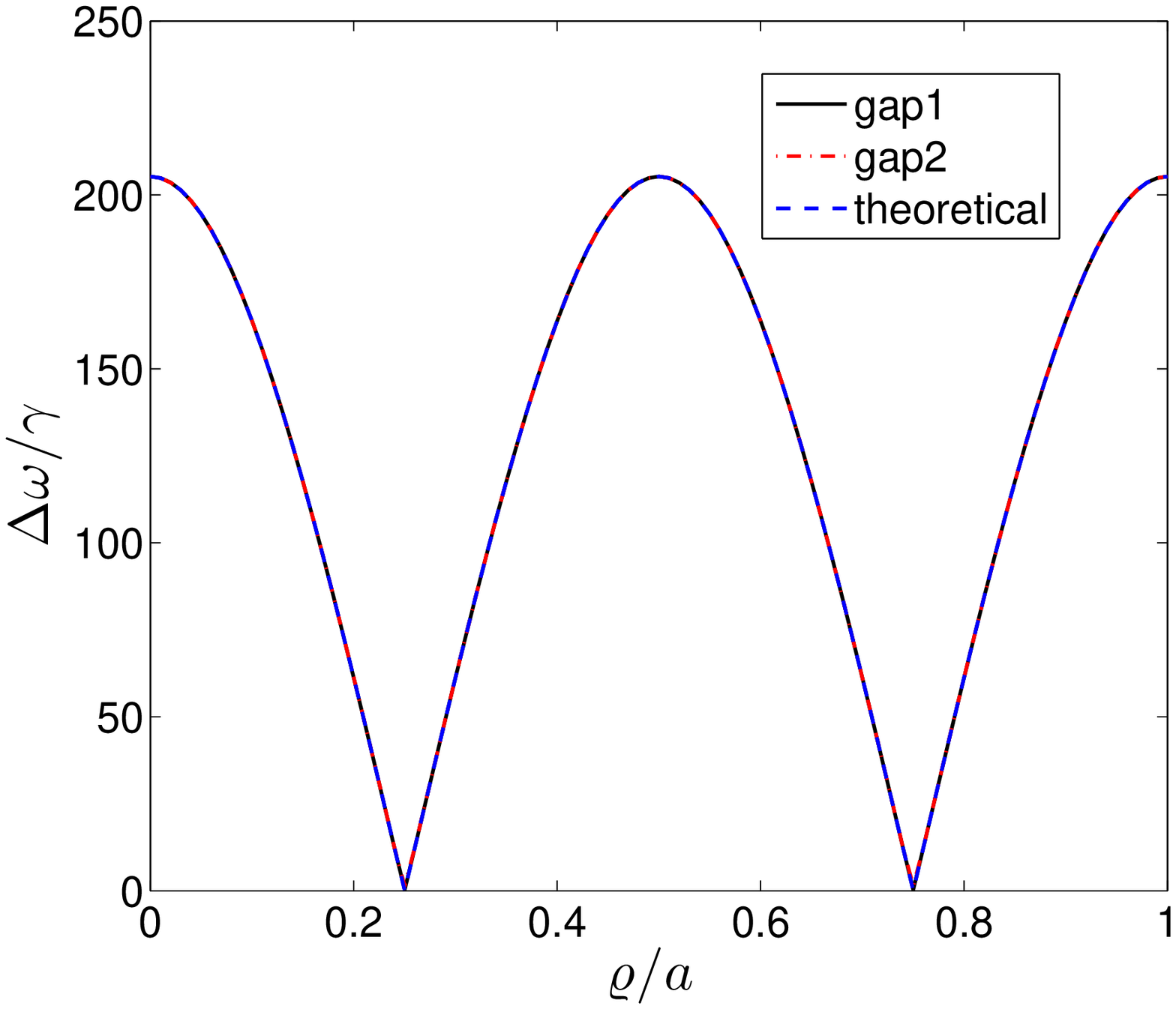}}
\vspace{.1in} \caption{\label{Fig:Spectrum:2} Same as in
Fig.~\ref{Fig:Spectrum:1} but for
$\omega_1=\omega_2=\omega_0-530 \gamma$. } \end{figure*}

\begin{figure*} \subfigure[]{
\includegraphics[width=.4\textwidth]{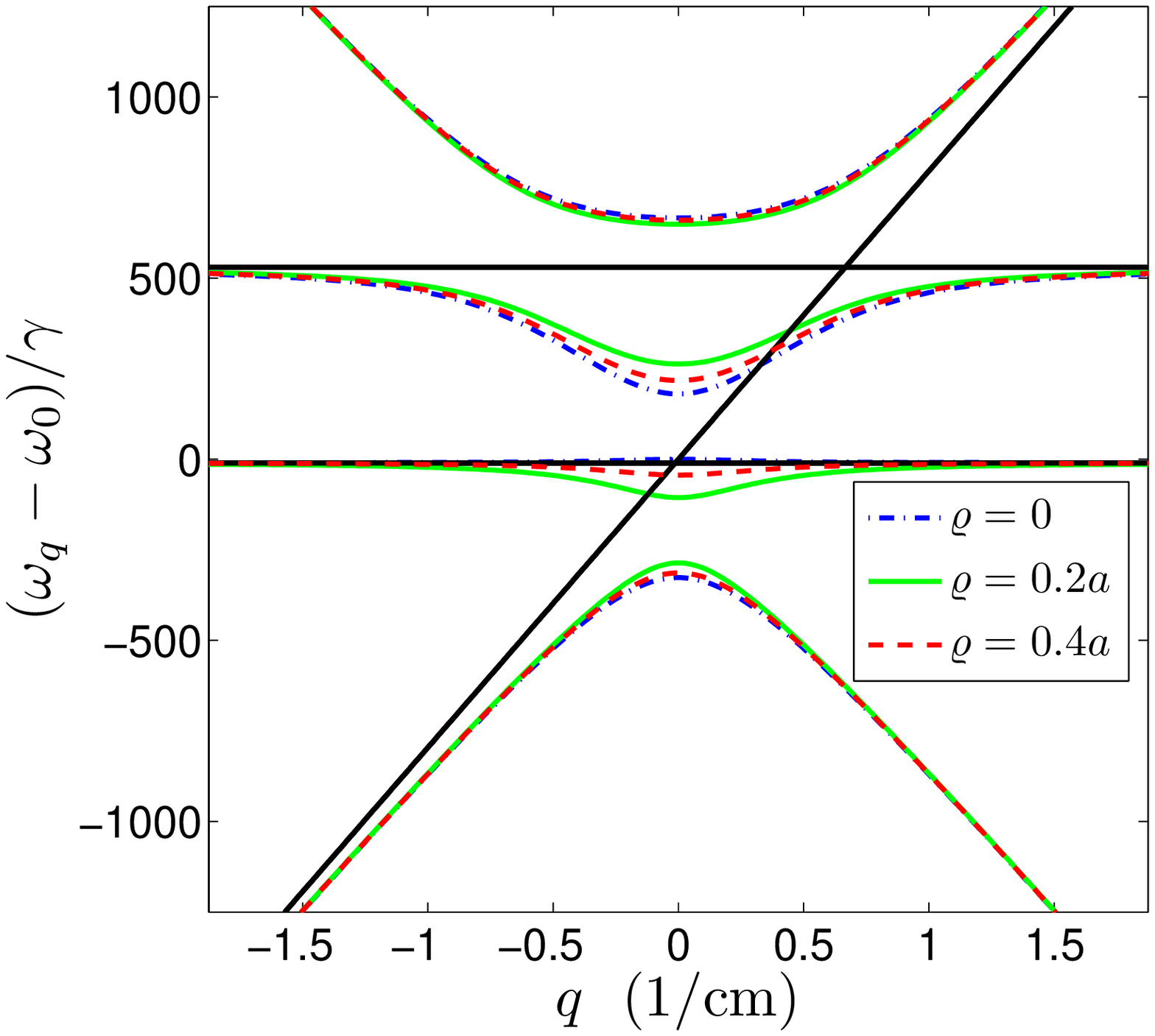}}
\subfigure[]{
\includegraphics[width=.4\textwidth]{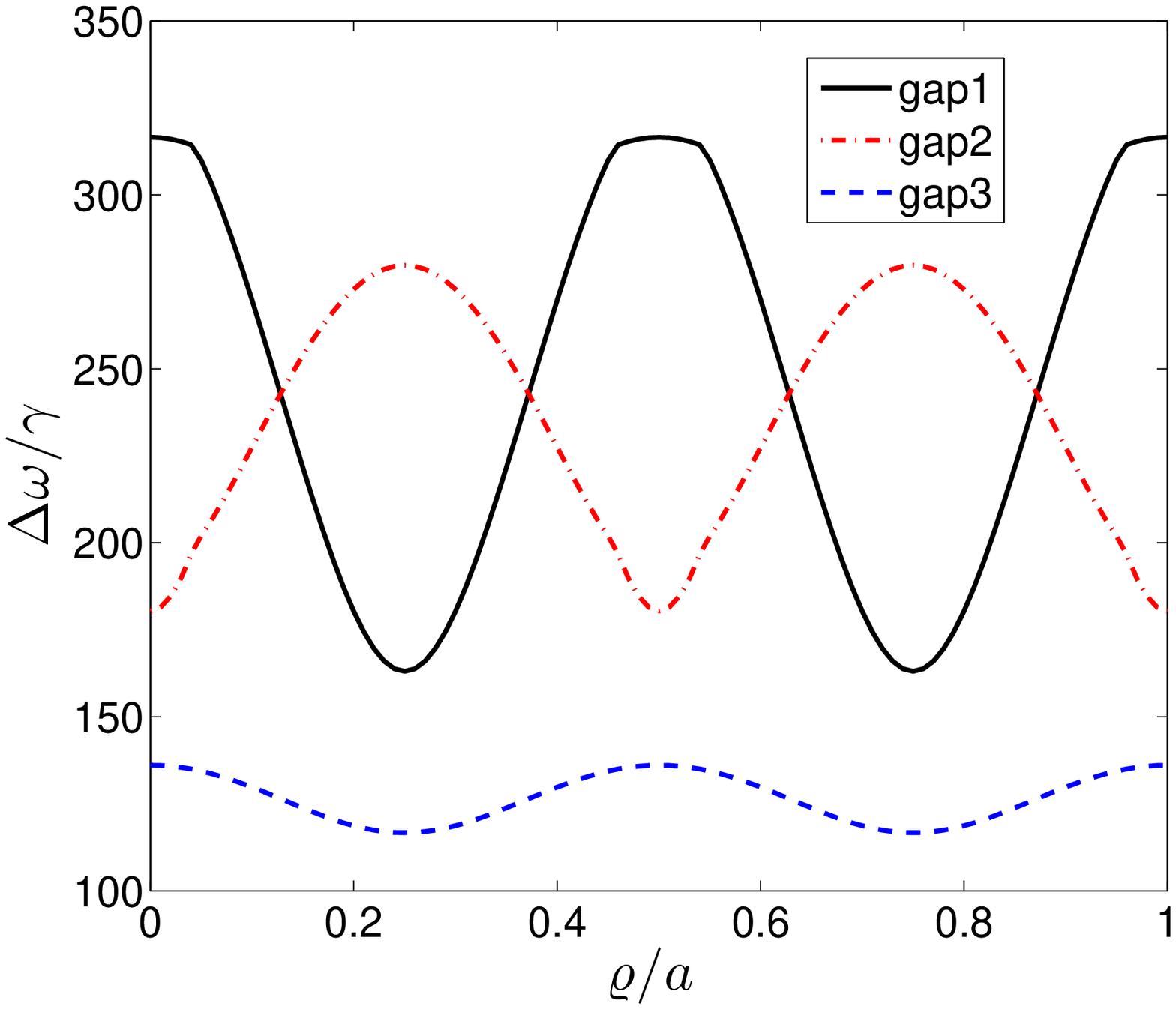}}
\vspace{.1in} \caption{\label{Fig:Spectrum:3} Same as in
Fig.~\ref{Fig:Spectrum:1} but for $\omega_1=\omega_0-10\gamma$ and
$\omega_2=\omega_0+530\gamma$. } \end{figure*}

\begin{figure*} \subfigure[]{
\includegraphics[width=.4\textwidth]{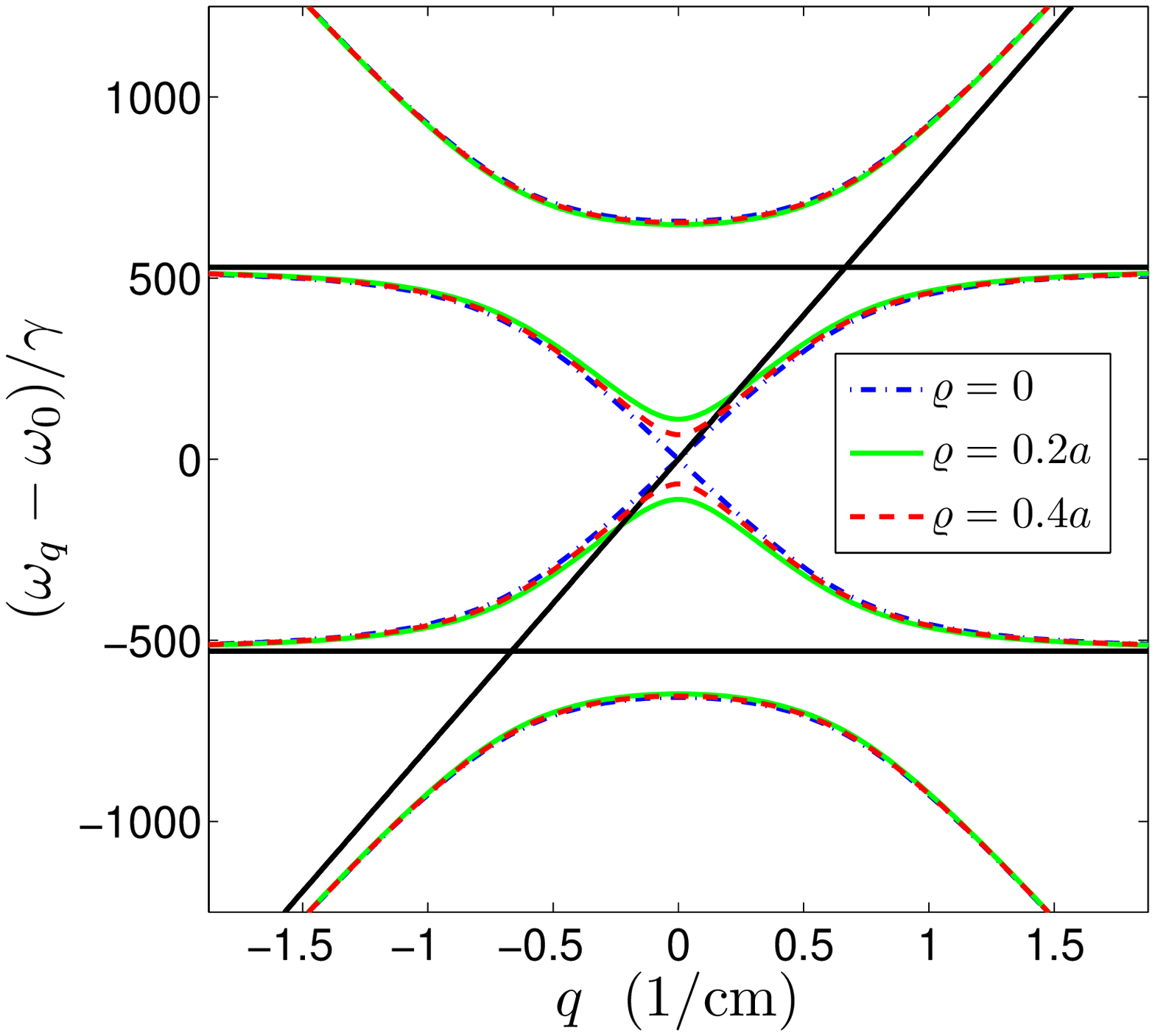}}
\vspace{.1in} \subfigure[]{
\includegraphics[width=.4\textwidth]{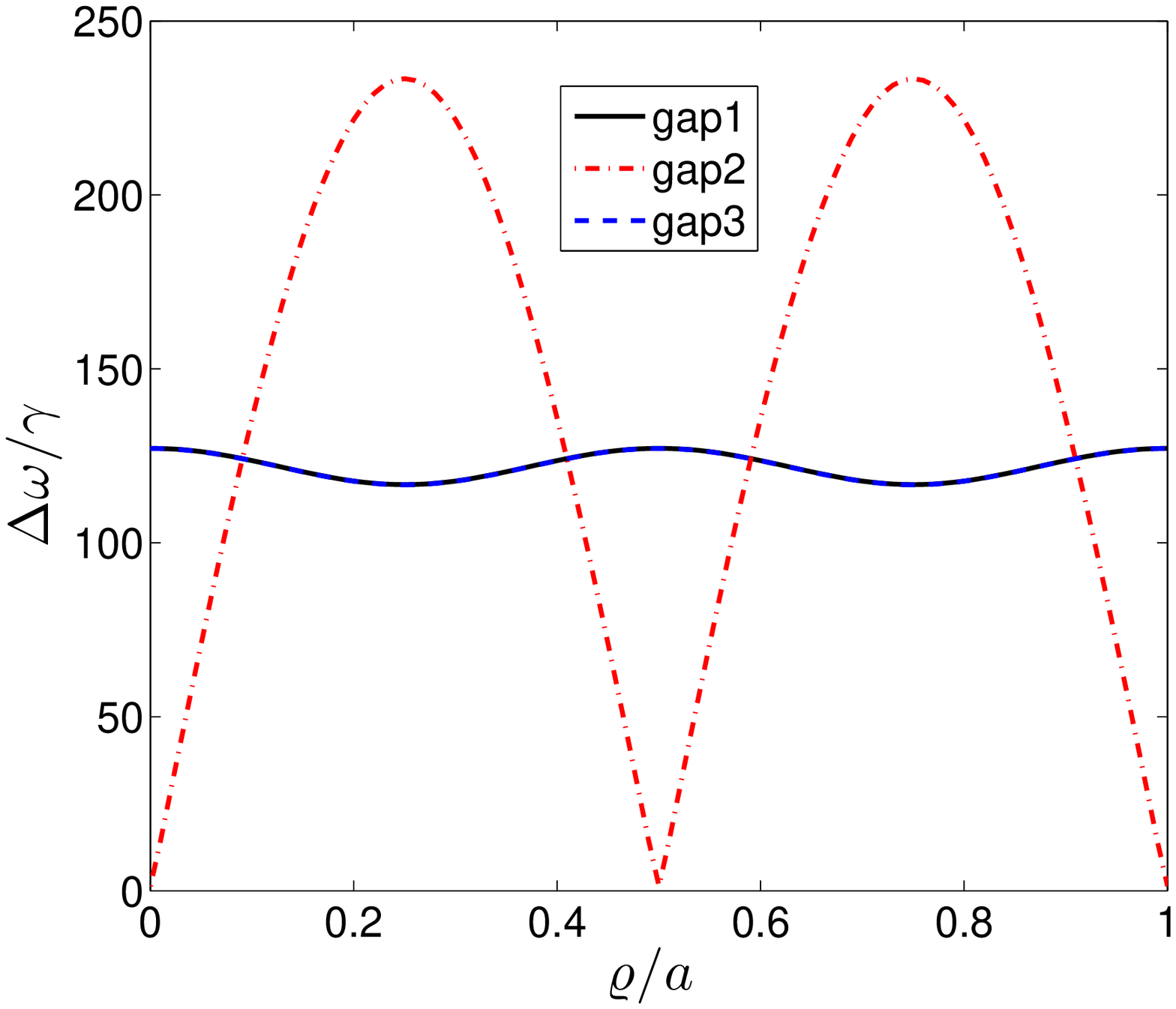}}
\vspace{.1in} \caption{\label{Fig:Spectrum:4}  Same as in
Fig.~\ref{Fig:Spectrum:1} but for $\omega_1=\omega_0-530\gamma$
and $\omega_2=\omega_0+530\gamma$. Note that gap $1$ and $3$ overlap. } \end{figure*}

These analytical results, obtained in a specific parameter
regimes, are confirmed by the results of the numerical spectra,
which are evaluated from Eqs.~(\ref{eqn:bandstructure}) by summing
over 40 BZs. The photonic spectra are shown in
Figs.~\ref{Fig:Spectrum:1}-\ref{Fig:Spectrum:2}(a), where the
polariton dispersion relation is reported around $q\simeq 0$ for
$\varrho=0,0.2a,0.4 a$ for several values of $\omega_1$, setting
$\omega_1=\omega_2$. The size of the bandgaps $\Delta\omega_+=
\nu_{2,+}-\nu_{1,+}$ and $\Delta\omega_-=\nu_{2,-}-\nu_{1,-}$ as a
function of $\varrho$ are displayed in 
Figs.~\ref{Fig:Spectrum:1}-\ref{Fig:Spectrum:2}(b), showing that
the size of the gap is controlled by $\varrho$, and it vanishes at 
Figures~\ref{Fig:Spectrum:3} and~\ref{Fig:Spectrum:4} display the
photonic spectra when the atoms composing the Wigner-Seitz cell
are of different species, in the case in which both interact with
the probe but the resonance frequency of the respective dipolar
transition is different. Figure~\ref{Fig:Spectrum:3} displays the
photonic spectrum for the specific case in which one atomic
transition is quasi-resonant, while the second is far detuned. In
this case three photonic bandgaps appear, which vary largely as a
function of $\varrho$, and in such a way that while one is
minimum, the other two are maximum, and vice versa.
Figure~\ref{Fig:Spectrum:4} displays the case in which the two
atoms composing the cell are far detuned from the probe, with
detunings of opposite signs. The spectrum is also characterized by
three bandgaps.

In this treatment we neglected atomic absorption, however the
evaluated bandgaps are significantly larger than the linewidth
$\gamma$, and hence they can be experimentally observed. The
effect of absorption is considered in the following section in the
framework of a semiclassical model, where also finite size effects are accounted for.


\section{Probe transmission spectra} \label{Sec:3}

We now study the response of the biperiodic array of atoms to an
external probe, and evaluate probe reflection and transmission as
a function of the probe frequency considering the finite size effects. In this section we assume a three-dimensional array of atoms, where $N$ lattice planes are aligned along the $x$ direction.

The coupling of an external
probe to the atomic medium is described by the Maxwell equations
for the electric field, in the presence of a dielectric in the
region of space $0\le x\le L$, with $L=Ma$. The equations for a
probe at frequency $\omega_p$ and wave vector $k_p=\omega_p/c$,
propagating along $x$, read \begin{equation} \label{Maxwell}
[\partial^2_x+k_p^2]{\bf E}=-4\pi k_p^2{\bf P}\end{equation} where
${\bf P}(x,t)$ is the medium polarization. Continuity of the field
and its spatial derivative at the interfaces imposes
\begin{equation}\begin{array}{l}
{\bf E}(x=0_+)={\bf E}(x=0_-)\,,\\
\partial_z {\bf E}(x=0_-)-\partial_z {\bf E}(x=0_+)=4\pi k_p^2{\bf
P}(x=0)\,,\end{array} \label{boundary} \end{equation} and
similarly at $x=L$. We consider the classical limit of these
equations, which is found from the quantum model within an
input-output formalism when the probe is a coherent
state~\cite{Carusotto}. In particular, we calculate the
reflection and transmission coefficients of an incident field by
means of the transfer matrix method.

We consider a probe field propagating at normal incidence at $x=0$
to a single polarizable plane composed by atoms at resonance
frequency $\omega_j$ and linewidth $\gamma_j$. In the linear
response regime, ${\bf P}=n_s\alpha_j\delta(x){\bf E}$ in
Eqs.~(\ref{Maxwell}) and (\ref{boundary}), where $n_s$ is the
surface particle density and \begin{equation}
\alpha_j=\dfrac{3}{4\pi^2}\dfrac{\varepsilon_0}{\hbar}\lambda_p^3
\left(\frac{2\delta_j/\gamma_j+i}{1+4\delta_j^2/\gamma_j^2}\right)
\end{equation} is the classical polarizability per particle, with
$\delta_j=\omega_j-\omega_p$ the detuning with respect to the atomic
transition~\cite{Deutsch1995a}. Denoting by ${\bf E_0}$ the amplitude of the incident
field, we write the reflected and transmitted fields as ${\bf
E}_r=r_j {\bf E}_0e^{-ik_px}$ and ${\bf E}_t=t_j{\bf
E}_0e^{ik_px}$, where $r_j$ and $t_j$ are the transmission
coefficients at a plane with polarizability $\alpha_j$, and
read~\cite{Deutsch1995a} \begin{equation}
r_j=\frac{i\xi_j}{1-i\xi_j},\;\;\;t_j=\frac{1}{1-i\xi_j},
\end{equation} with $\xi_j=2\pi k_pn_s\alpha_j$. The transfer
matrix ${\mathcal M}_j$ relates forward- and backward-traveling
waves ${\bf E}^+_t$, ${\bf E}^-_t$  on the right-hand side of a plane
to those on the left-hand side ${\bf E}^+_r$, ${\bf E}^-_r$,
according to the relation \begin{equation} \left(\begin{array}{l}
{\bf E}^+_t\\{\bf E}^-_t\end{array}\right)=\mathcal{M}_j
\left(\begin{array}{l} {\bf E}^+_r\\{\bf
E}^-_r\end{array}\right)\,, \end{equation} thereby automatically
accounting for all interference effects accumulated along the way.
The transfer matrix for a single period is given by the product of
the transfer matrix across the boundary of an atomic plane and
free propagation for a distance $d_j$ \begin{equation}
\mathcal{M}_{d_j}=\frac{1}{t_j}\left(\begin{array}{cc}
t^2_j-r^2_j&r_j\\-r_j&1\end{array}\right)
\left(\begin{array}{cc}e^{ik_pd_j}&0\\
0&e^{-ik_pd_j}\end{array}\right)\,. \label{tm} \end{equation} For
any transfer matrix
$\tilde{\mathcal{M}}=\prod_{j}\mathcal{M}_{d_j}$, generated by the
product of single-period transfer matrices, the reflection and the
transmission coefficients associated with the matrix elements are
\begin{equation}
r=\frac{\tilde{\mathcal{M}}_{12}}{\tilde{\mathcal{M}}_{22}}\,,\;\;\;
t=\frac{1}{\tilde{\mathcal{M}}_{22}}\,, \end{equation} as is
easily verified for the special case of matrix in Eq. (\ref{tm}).
The transmission and reflection coefficients for the biperiodic
lattice of $M=N/2$ planes is found as a function of the transfer
matrix
$\mathcal{M}_{d_1d_2}=\mathcal{M}_{d_1}\cdot\mathcal{M}_{d_2}$ for
a single dimer, with $d_1=\varrho$, $d_2=a-\varrho$. In
particular, the transmission coefficient reads \begin{widetext}
\begin{equation} t_n=\frac{1}{(\mathcal{M}_{d_1d_2}^n)_{22}}=
\frac{\sin{\Theta}}{\sin\Theta\cos(n\Theta)+i\sin(n\Theta)[\xi_1\xi_2\sin
k_p\varrho- (1-\xi_1\xi_2)\sin k_pa-(\xi_1+\xi_2)\cos k_pa]}\,,
\label{tcoeffdim} \end{equation} with $\Theta={\rm
acos}[\xi_1\xi_2\cos k_p\varrho+ (1-\xi_1\xi_2)\cos
k_pa-(\xi_1+\xi_2)\sin k_pa]$. The elementary cell dephasing
$\Theta$ can be written as function of the single slice dephasings
$\Theta_{1,2}={\rm acos}(\cos k_pd_{1,2}-\xi_{1,2}\sin
k_pd_{1,2})$ as following \begin{equation} \Theta={\rm
acos}\left[\cos(\Theta_1+\Theta_2)+\sin\Theta_1\sin\Theta_2-\sqrt{
\sin^2\Theta_1\sin^2\Theta_2-\xi_1\xi_2\sin^2k_p\varrho}\right],
\end{equation} showing that the term $\sin^2k_p\varrho$ governs
the difference between a simple addition of the dephasings
$\Theta_j$, as already pointed out in the calculation of the
spectrum.

If ${\rm Im}\Theta>{\rm Re} \Theta$, which occurs around
$\omega_p\simeq\omega_1,\omega_2$, in the limit
$n\rightarrow\infty$ the transmission coefficient
(\ref{tcoeffdim}) can be approximated by \begin{equation}
\lim_{n\rightarrow\infty}\!t_n\!=\! \frac{2e^{-n{\rm
Im}\Theta}\sin{\Theta}}{\sin{\Theta}\!+\!\xi_1\xi_2\sin
k_p\varrho\!-\! (1-\xi_1\xi_2)\sin k_pa\!-\!(\xi_1+\xi_2)\cos
k_pa}. \end{equation} \end{widetext} For vanishing values of ${\rm
Im}\Theta$, the effect of multiple scattering cannot be factorized
in a simple attenuation coefficient, and gives rise to
interference structures as a function of $\varrho$.

\begin{figure} \begin{center}
\includegraphics[width=0.49\linewidth,clip=true]{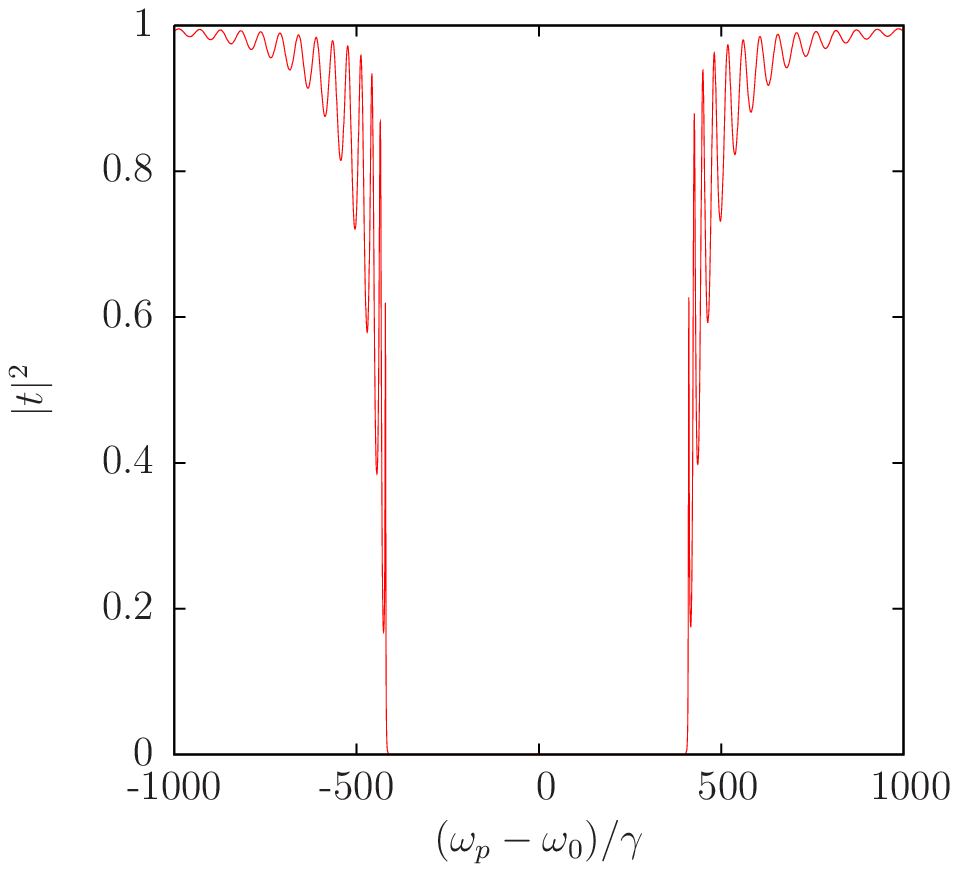}
\includegraphics[width=0.49\linewidth,clip=true]{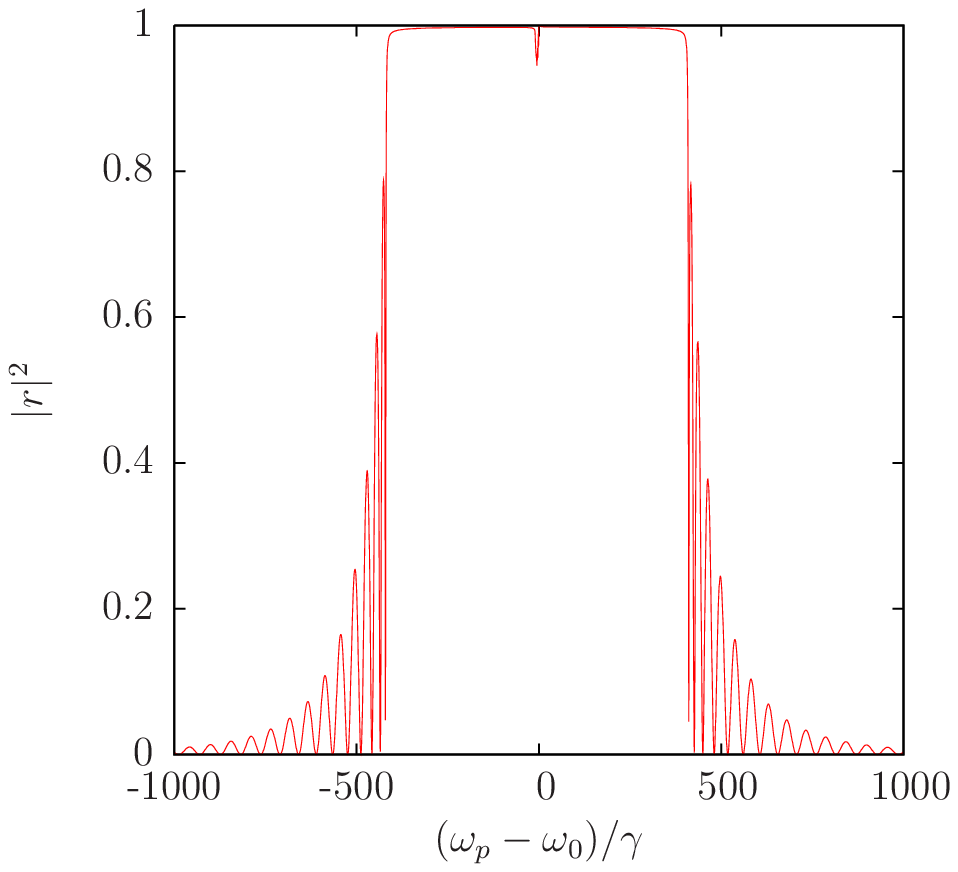}
\caption{\label{fig1tr} The absolute square of the transmission
coefficient (left panel) and of the reflection coefficient (right
panel) as a function of $(\omega_p-\omega)/\gamma$ is plotted for
a lattice of $^{85}$Rb atoms with areal density $n_s=5.7\cdot
10^{-2}\mu$m$^{-2}$, trapped by laser beams detuned to the blue of
resonance by $10\gamma$ and for $10^6$ atomic planes.}
\end{center} \end{figure}

In order to compare the transmission and reflection spectra
with the results obtained for the photonic band structure,
we consider a cloud of $^{85}$Rb atoms in a
quasi-one-dimensional geometry, confined in a very
long lattice, taking $N=10^6$ atomic
slices.
We fix the areal density at $n_s=5.7\cdot 10^{-2}$
$\mu$m$^2$, and we consider that the atoms are in the same
internal state, so that $\omega_2=\omega_1$ and $\gamma_2=\gamma_1$. In this
section we will refer to optical lattices generated by laser beams
with a frequency $\omega_0=\omega_1+10\gamma$, where
$\omega_1$ and $\gamma_1$ refer to the $D2$ atomic resonance at
$\lambda=780$ nm. Figure~\ref{fig1tr} displays the absolute square of the
transmission coefficient (left panel) and of the reflection
coefficient (right panel) of a probe beam traveling through a
monoperiodic sequence of atomic planes ($\varrho=0$).
For our choice of the parameters we find
the edges of the gap at $\pm 420\gamma$. The width of the gap
depends on our choice for $n_s$. The spectrum of transmission agrees qualitatively, as a function of $\varrho$, with the one shown in Fig.~\ref{Fig:Spectrum:1}, where a strictly one-dimensional system was considered. Notice that the small peak in the transmission and the
small dip in the reflectivity at the atomic frequency are a mark
of scattering losses.

\begin{figure} \begin{center}
\includegraphics[width=0.49\linewidth,clip=true]{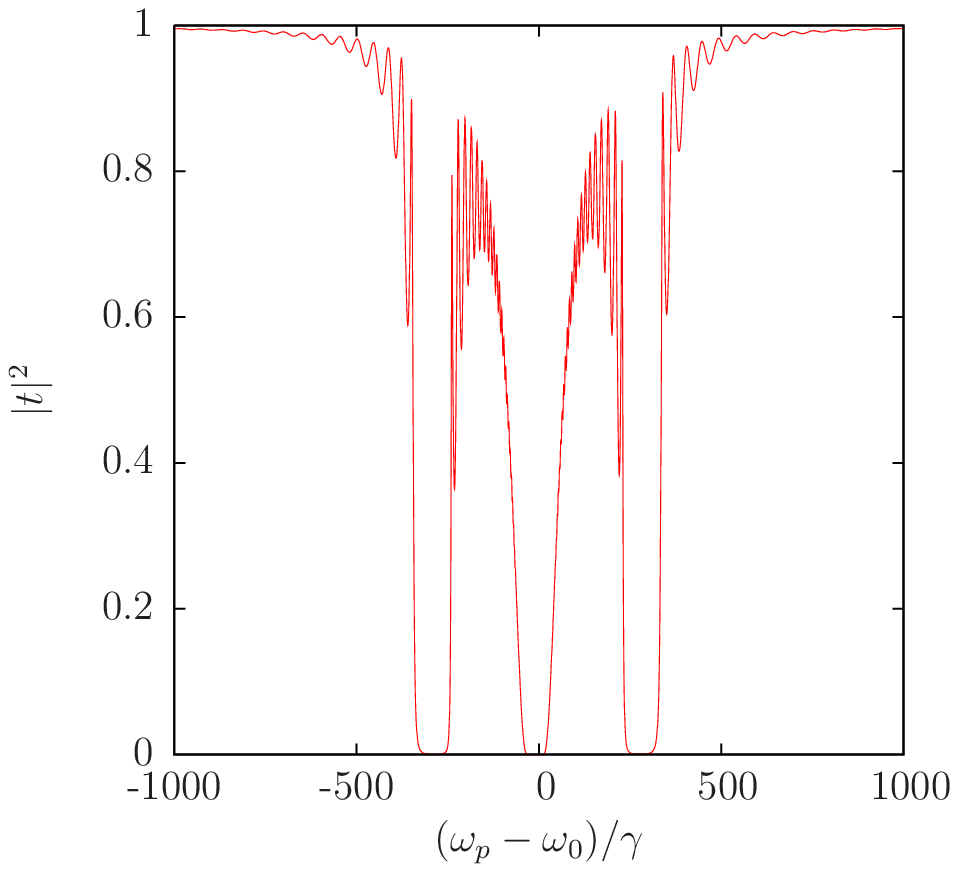}
\includegraphics[width=0.49\linewidth,clip=true]{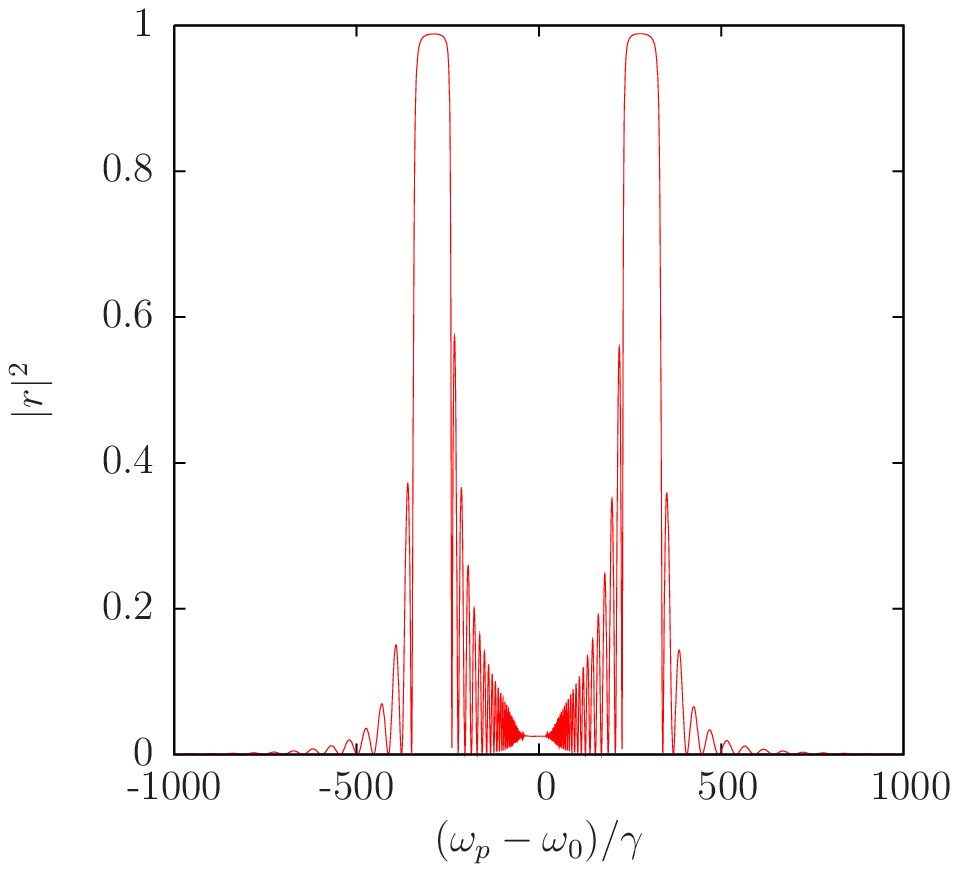}
\caption{\label{fig2tr}Same as Fig. \ref{fig1tr}, but
$\varrho=0.2\,\lambda$.} \end{center} \end{figure}

\begin{figure}
\begin{center}
\includegraphics[width=0.49\linewidth,clip=true]{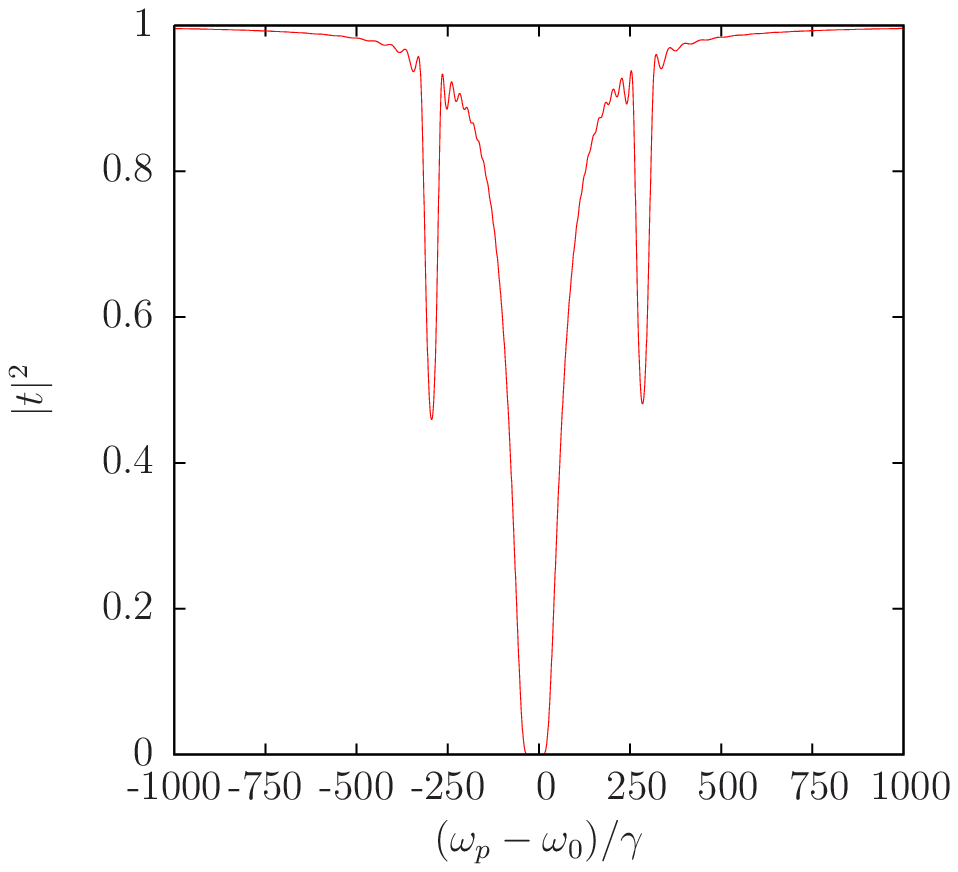}
\includegraphics[width=0.49\linewidth,clip=true]{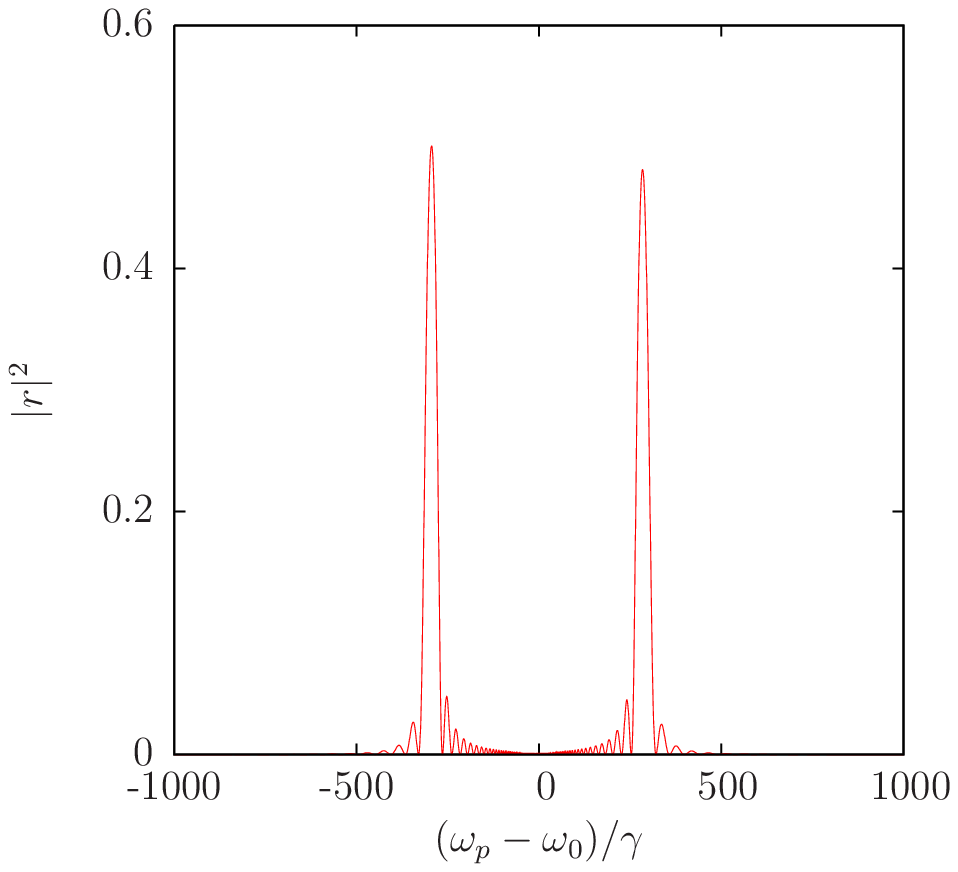}
\caption{\label{fig3tr}Same as Fig. \ref{fig1tr}, but for
$\varrho=0.24\,\lambda$.}
\end{center} \end{figure}

In a bichromatic lattice, such as that shown in
Fig.~\ref{lattice}, the structure of the transmission and
reflection spectrum is deeply modified by the multiperiodicity.
Figures \ref{fig2tr} and \ref{fig3tr} refer to the cases
$\varrho=0.2a$, and $0.24 a$ respectively.

In agreement with the results for the band structure at $\varrho=0.2 a$ in Fig.~\ref{Fig:Spectrum:1}, at the
center of the gap it appears a high transmission region divided in two symmetric
parts by the absorption line (Fig. \ref{fig2tr}). By increasing $\varrho$
(Fig. \ref{fig3tr}), the width of the mini-band increases  and the gaps become thinner. At $\varrho=0.25 a$, the gaps close and the atomic lattice becomes
transparent in this region of frequencies, everywhere except for
$\omega_p\simeq\omega_0$, because of the absorption. For this particular
value of $\varrho$, ${\mathcal M}_{d_2}\simeq{\mathcal M}_{-d_1}\simeq{\mathcal M}_{d_1}^{-1}$, if $\xi_2=\xi_1$ and neglecting absorption (the imaginary part of $\xi_j$).

\section{Bichromatic lattice inside a single-mode cavity}\label{Sec:4}

The observation of sufficiently large photonic bandgaps in free
space requires a large number of lattice planes in a well
controlled periodic structure, which is experimentally
challenging. Nevertheless, observable effects can be found in
small systems when coupling the atomic transition, for instance, to the modes of a fiber~\cite{Ketterle} or of
an optical resonator~\cite{Zimmermann_Bragg}.

Let us assume that the dipolar transition of the atoms couples
strongly with the single mode of a standing-wave optical
resonator, which probes the system. 
The Hamiltonian for the dynamics inside the cavity is now
$H'=H_{\rm dip}+H_c+H_{\rm int}'$, where $H_{\rm dip}$ is given in
Eq.~(\ref{H:dip}),\begin{eqnarray} H_c=\hbar\omega_ca^{\dagger}a
\end{eqnarray} describes the cavity field at frequency $\omega_c$,
with $a$, $a^{\dagger}$ annihilation and creation operators of a
cavity photon, and \begin{eqnarray} H_{\rm int}'=\hbar
\sum_jg_j\cos(kx_j+\varphi)\sigma_j^{\dagger}a+{\rm H.c}
\end{eqnarray} is the Jaynes-Cummings Hamiltonian, with $k$ the
cavity mode wave vector and $g_j$ the coupling strength, which
depends on the dipolar moment $\mathcal D_j$ of the atom at
position $x_j$. As a function of the cavity parameters,
$g_j=\sqrt{\varsigma/(4\pi A)}\sqrt{\gamma\delta\omega}$, with
$\varsigma$ the scattering cross section in free space, $A=\pi w_c^2/4$ with $w_c$ the
cavity mode waist, and $\delta\omega=2\pi c/L$ the free spectral
range, with $L$ the cavity length~\cite{Kimble_Review}. The phase
$\varphi$ accounts for the dephasing between the cavity mode
lattice and the atomic lattice.

The system is probed by an external weak coherent pump at
intensity $\eta$ and frequency $\omega_p$, which is coupled to the
resonator. In this limit, we make the Holstein-Primakoff
transformation and keep only the linear term. The resulting
Heisenberg-Langevin equations of motion for cavity mode and spin
wave operators read~\cite{Walls} \begin{subequations}
\label{QLE:123} \begin{eqnarray} \label{QLE:1}\dot{a}&=& -{\rm
i}\delta_c a-\kappa a+\eta+\sqrt{2\kappa}\zeta(t)-{\rm i}
\frac{\sqrt{M}}{2}
\nonumber\\
&\times& \sum_{Q\in BZ,Q=k-G}
\Bigl[g_1 \left({\rm e}^{-{\rm i} \varphi} b_Q+ {\rm
e}^{{\rm i} \varphi} b_{-Q}\right)\,,\nonumber\\
&+&g_2\left({\rm e}^{-{\rm
i}(k\varrho+\varphi)}d_Q+{\rm e}^{{\rm
i}(k\varrho+\varphi)}d_{-Q}\right)\Bigr]\\
\dot{b}_{\pm Q}&=&-\left({\rm i}\delta_1
+\frac{\gamma_1}{2}\right)b_{\pm Q}-\frac{{\rm i}}{2} \sqrt{M}
g_{1} {\rm e}^{\pm{\rm i}\varphi} a
+\sqrt{\gamma_1}\mathcal B_{1,Q}\,,\nonumber\\
&&\\
\label{QLE:3}\dot{d}_{\pm Q}&=&-\left({\rm i}\delta_2
+\frac{\gamma_2}{2}\right) d_{\pm Q}-\frac{\rm i}{2} \sqrt{M}
g_{2} {\rm e}^{\pm {\rm i}(k\varrho+\varphi)}
a+\sqrt{\gamma_2}\mathcal B_{2,Q}\,,\nonumber\\
\end{eqnarray}\end{subequations} where
$\delta_c=\omega_c-\omega_p$ and $\kappa$ is the cavity linewidth. The noise operators
$\zeta(t)$, $\mathcal B_{j,Q}$, have zero mean value and satisfy
the relation $\langle \zeta(t)\zeta(t')^{\dagger}\rangle =\langle
\mathcal B_{j,Q}(t)\mathcal
B_{j,Q}(t')^{\dagger}\rangle=\delta(t-t')$ (we assume the
electromagnetic field in the vacuum).

Equations~(\ref{QLE:1})-(\ref{QLE:3}) describe the coupling
between the cavity mode, at momentum $k$, and the spin waves at
quasi-momentum $Q$ (inside the first Brillouin zone), such that
$Q+G=k$ where $G$ is a vector of the reciprocal lattice. We
identify two relevant cases, when (i) $k\neq\mathcal N \pi/a$, and
(ii)  $k=\mathcal N \pi/a$, where $\mathcal N$ is an integer.

For $k\neq\mathcal N \pi/a$ the system, composed by cavity
potential and bichromatic lattice, is not periodic. The
eigenfrequencies of the homogeneous equations can be simply found
for the case $\delta_1=\delta_2=\Delta$,
$\gamma_1=\gamma_2=\gamma$, and read \begin{eqnarray}
\nu_0&=&\Delta-{\rm i}\gamma/2\,, \label{nu:0} \\
\nu_{\pm}&=&\frac{\delta_c+\Delta-{\rm i}(\kappa+\gamma/2)}{2}\label{nu:pm}\\
&&\pm  \sqrt{\frac{1}{4}\left(\delta_c-\Delta-{\rm i}\kappa+{\rm
i}\frac{\gamma}{2} \right)^2+M\mathcal R}\,, \nonumber
\end{eqnarray} where \begin{equation}\mathcal
R=(g_1^2+g_2^2)/2\label{R:0}.\end{equation} Here, the real part
gives the position of the resonances, while the imaginary part
gives the corresponding linewidth. The eigenmodes at frequency
$\nu_0$ are pure spin waves, and hence correspond to collective
dipolar excitations which are decoupled from the cavity field. The
eigenmodes at frequency $\nu_{\pm}$ are polariton excitations. We
remark that the frequencies $\nu_{\pm}$ do not depend on
$\varrho$.

For $k=\mathcal N \frac{\pi}{a}$, the system, composed by cavity
potential and bichromatic lattice, is periodic. The cavity mode
couples to the spin waves $Q=0$ or $\pi/a$, depending on whether $\mathcal N$ is even or odd, respectively. The
eigenfrequencies of the homogeneous equations (for the specific
case $\delta_1=\delta_2=\Delta$, $\gamma_1=\gamma_2=\gamma$) have
the same form as in Eqs.~(\ref{nu:0}-\ref{nu:pm}), whereby now in
Eq.~(\ref{nu:pm}) the coefficient $\mathcal R$ reads
\begin{equation} \label{R:1}\mathcal R= g_1^2\cos^2 \varphi+g_2^2
\cos^2(k\varrho+\varphi)\,.\end{equation} We note that the
eigenfrequencies in this case explicitly depend on $\varrho$. This
result is also found when the bichromatic lattice is replaced by
one single cell, by rescaling the coupling strength of each atom
inside the cell as $g_{{\rm eff},j}=\sqrt{M} g_{j}$.

\begin{figure} \begin{center}
\includegraphics[width=0.49\linewidth,clip=true]{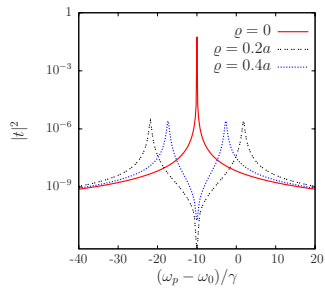}
\caption{\label{Fig:cav:1}(Color online) The absolute square of the transmission
coefficient as a function of $(\omega_p-\omega_0)$ (in units of
$\gamma$) is plotted for a lattice of $^{85}$Rb atoms with areal
density $n_s=5.7\cdot 10^{-2}$ $\mu$m$^{-2}$, trapped by laser
beams detuned to the blue of resonance by 10 $\gamma$, for $200$
planes, in the presence of a cavity of length $L=85$ mm and
finesse $\mathcal{F}=170000$, and for $\varphi=\pi/2$. Note that
for $\varrho=0$ the atoms are trapped at the nodes of the
resonator and hence do not interact with the cavity field.}
\end{center} \end{figure}

\begin{figure} \begin{center}
\includegraphics[width=0.49\linewidth,clip=true]{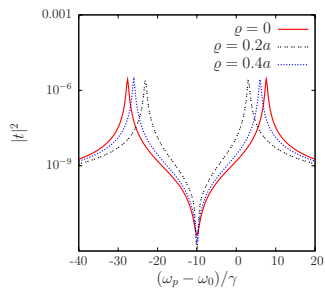}
\caption{\label{Fig:cav:2} Same as Fig.~\ref{Fig:cav:1}, but for
$\varphi=0$.}
 \end{center} \end{figure}

We now discuss the intensity of the field at the cavity output as
a function of the probe frequency $\omega_p$ for various
parameters, by solving Eqs.~(\ref{QLE:1})-(\ref{QLE:3})
numerically. The quantity we study is the number of photons per
unit time $I(\omega_p)=\langle a_{\rm out}^{\dagger}a_{\rm
out}\rangle$, where $a_{\rm out}$ is the field at the cavity
output, $a_{\rm out}=\sqrt{2\kappa}a-\zeta$, and the average is
taken over the state of the system and the vacuum state of the
e.m.-field outside the resonator~\cite{Walls}. Hence, we find
\begin{equation} I(\omega_p)=2\kappa \langle a^{\dagger}a\rangle
=\frac{2\kappa \eta^2
(\Delta^2+\gamma^2/4)}{(\kappa\gamma/2-\delta_c\Delta+M\mathcal
R)^2+(\Delta \kappa+\delta_c\gamma/2)^2 } \,,\label{eqn:Iout}
\end{equation} where we have used the steady state solution of
Eq.~(\ref{QLE:1}). In the strong coupling regime, when the
cooperativity $\mathcal C\sim M\mathcal R/2\kappa\gamma\gg
1$~\cite{Kimble_Review}, the intensity $I(\omega_p)$ at the cavity
output exhibits two well defined maxima at the frequencies
\begin{equation} \omega_p^0=\frac{\omega_c+\omega_a}{2}\pm
\sqrt{\left(\frac{\omega_c-\omega_a}{2}\right)^2+M\mathcal R}
\label{Rabispli2} \end{equation} and corresponding to the vacuum
Rabi splitting for this system~\cite{Kimble_Review}. This can also
be seen in Figs. \ref{Fig:cav:1} and \ref{Fig:cav:2}. It is
interesting to note that for $\Delta=0$ and large cooperativity
the cavity output field goes to zero as $1/\mathcal C^2$. This is an
interference effect, where the atomic polarization inside the
cavity form a field equal and opposite to the driving pump, such
that the cavity field is effectively empty. Energy is in this case
dissipated by the atoms. This behaviour has been first predicted
in~\cite{CIT} under the name "cavity induced transparency".

We evaluate the cavity transmission spectrum using the parameters
of the setup in~\cite{Kruse2003}, and consider $^{85}$Rb atoms
inside a resonator with length $L=85$mm, finesse $\mathcal
F=170000$, loss rate $\kappa=2\pi\times 21$ KHz, beam waist
$w=130~\mu$m, and an average occupation per site $\bar n=3000$,
that corresponds to an areal density $n_s\simeq 5.7\cdot
10^{-2}\mu$m$^{-2}$. The reflectivity of the cavity mirrors is
$|r|^2\simeq 1- 1.8\times 10^{-5}$. The transmission spectrum is
calculated assuming that $N=200$ planes ($M=100$) are confined 
inside the resonator. The positions of the peaks correspond to
those predicted by Eq.~(\ref{Rabispli2})
by taking into account multiple occupancy of the lattice sites
rescaling the coupling strengths as $g_j\rightarrow\sqrt{\bar
n}g_j$. Figure~\ref{Fig:cav:1} displays the squared transmission
as a function of the probe frequency for the case $\varphi=\pi/2$
and the values $\varrho=0,0.2 a,0.4 a$. Note that the case $\varrho=0$
corresponds to all atoms at the antinodes of the resonator, and it
is hence equivalent to the situation in which the cavity is
empty. Figure~\ref{Fig:cav:2} displays the transmission spectrum
is the optical lattice trapping the atom is shifted so that
$\varphi=0$, showing that the form changes substantially. By
varying $\varphi$, hence, information on the interparticle
distance in the Wigner-Seitz cell can be gained. The minimum at
$\omega_p=\omega_0$, corresponding to the cavity induced transparency
behaviour, is here visible.
 
\section{Conclusion} \label{Sec:5}

The photonic properties of biperiodic optical lattices are
critically determined by the interparticle distance $\varrho$
inside the primitive Wigner-Seitz cell. We have derived a model
describing light propagation for a weak probe, and its response to
probe propagation in free space and inside of a cavity. We have
found that, depending on $\varrho$, in free space the system may
or may not exhibit photonic bandgaps about the atomic frequency.
This is a peculiar property, which makes the biperiodic crystal
different from the monoperiodic one, always exhibiting a bandgap
at the atomic frequency. In case there are photonic bandgaps
around this value, they occur in two or more ranges of frequencies. For
a finite crystal, relevant effects can be observed when the atoms
are confined inside an optical resonator. Here, the interparticle
distance $\varrho$ inside the primitive Wigner-Seitz cell
determines the properties of the transmission spectrum of a probe
at the cavity output.

Our study is based on a full quantum model for the light. In this
paper we have focused on the intensity of the transmitted and
reflected light, in the future we will study higher order
coherence of the scattered light. On the basis of studies made
with two atoms inside a cavity~\cite{SoniaPRA}, we expect that,
when considering saturation effects, the biperiodic optical
lattice can act as nonlinear-optical medium, whose properties may be
controlled by the interparticle distance $\varrho$. In the linear
response regime, it can be interesting to study higher order
coherence of the scattered light for various states of matter
inside the potential, with the aim of determining the quantum
state of the matter~\cite{Mekhov}. This may allow, in particular,
to detect experimentally novel states of matter realized in
bichromatic optical
lattices~\cite{Davidson2006,Fallani,Barmettler2008a}.
\section*{Acknowledgement} This work was carried out under the
HPC-EUROPA project (RII3-CT-2003-506079), with the support of the
European Community - Research Infrastructure Action under the FP6
"Structuring the European Research Area" Programme. Support by the
European Commission (EMALI, MRTN-CT-2006-035369; SCALA, Contract
No.\ 015714) and by the Spanish Ministerio de Educaci\'on y
Ciencia (Consolider Ingenio 2010 QOIT, CSD2006-00019; QNLP,
FIS2007-66944; Ramon-y-Cajal) are acknowledged.

 \end{document}